\documentclass[twocolumn,showpacs,preprintnumbers,amsmath,amssymb,aps,prl,superscriptaddress]{revtex4}
\usepackage{epsfig}
\usepackage{graphicx}
\usepackage{dcolumn}
\usepackage{bm}
\usepackage{color} 

\begin{document}

\title{
Electronic-Structure-Driven Magnetic Ordering in a Kondo Semiconductor CeOs$_2$Al$_{10}$.
}
\author{Shin-ichi Kimura}
\email{kimura@ims.ac.jp}
\affiliation{UVSOR Facility, Institute for Molecular Science, Okazaki 444-8585, Japan}
\affiliation{School of Physical Sciences, The Graduate University for Advanced Studies (SOKENDAI), Okazaki 444-8585, Japan}
\author{Takuya Iizuka}
\affiliation{School of Physical Sciences, The Graduate University for Advanced Studies (SOKENDAI), Okazaki 444-8585, Japan}
\author{Hidetoshi Miyazaki}
\affiliation{UVSOR Facility, Institute for Molecular Science, Okazaki 444-8585, Japan}
\author{Akinori Irizawa}
\altaffiliation{Present address: The Institute of Scientific and Industrial Research, Osaka University, Ibaraki, Osaka 567-0047, Japan}
\affiliation{Graduate School of Science and Technology, Kobe University, Nada-ku, Kobe 657-8501, Japan}
\author{Yuji Muro}
\affiliation{Department of Quantum Matter, ADSM, Hiroshima University, Higashi-Hiroshima, Hiroshima 739-8530, Japan}
\author{Toshiro Takabatake}
\affiliation{Department of Quantum Matter, ADSM, Hiroshima University, Higashi-Hiroshima, Hiroshima 739-8530, Japan}
\affiliation{Institute for Advanced Materials Research, Hiroshima University, Higashi-Hiroshima, Hiroshima 739-8530, Japan}
%
%
%
%
\date{\today}
\begin{abstract}
We report the anisotropic changes in the electronic structure of a Kondo semiconductor CeOs$_2$Al$_{10}$ across an anomalous antiferromagnetic ordering temperature ($T_0$) of 29 K, using optical conductivity spectra. 
The spectra along the $a$- and $c$-axes indicate that a $c$-$f$ hybridization gap emerges from a higher temperature continuously across $T_0$. 
Along the $b$-axis, on the other hand, a different energy gap with a peak at 20~meV appears below 39~K, which is higher temperature than $T_0$, because of structural distortion. 
The onset of the energy gap becomes visible below $T_0$. 
Our observation reveals that the electronic structure as well as the energy gap opening along the $b$-axis due to the structural distortion induces antiferromagnetic ordering below $T_0$.
\end{abstract}

%
%
%
%
\maketitle
%
%
Rare-earth intermetallic compounds provide useful materials with characteristic physical properties, such as heavy fermions, Kondo semiconductors/insulators, and so on, due to the interaction between the local $4f$ electrons and the conduction electrons; namely $c$-$f$ hybridization~\cite{Hewson1993}.
With regard to these properties, Kondo semiconductors are known to have a $c$-$f$ hybridization gap on the Fermi level ($E_{\rm F}$), even though the magnetic susceptibility at high temperature obeys the Curie-Weiss law, indicating the local $4f$ character~\cite{Takabatake1998}.
Typical Kondo semiconductors such as SmB$_6$, YbB$_{12}$, Ce$_3$Bi$_4$Pt$_3$, CeRhSb, and others studied previously have no phase transition below the Kondo temperature ($T_{\rm K}$) because the magnetic moments of Ce~$4f$ are quenched due to the Kondo singlet state via strong $c$-$f$ hybridization.

However, the recently discovered Kondo semiconductors Ce$M_2$Al$_{10}$ ($M$~=~Os, Ru) that show another anomalous phase transition at a characteristic temperature $T_0$ (28.7~K for $M$~=~Os, 27.3~K for Ru) below the Kondo temperature ($T_{\rm K}\sim$100~K for Os, 60~K for Ru) have been found~\cite{Strydom2009,Nishioka2009,Matsumura2009}, in contrast to the related material CeFe$_2$Al$_{10}$, which is a typical Kondo semiconductor even at very low temperatures~\cite{Muro2009,Muro2010-1}.
Below $T_0$, CeOs$_2$Al$_{10}$ shows a different semiconducting activation type electrical resistivity from that above $T_0$ despite the fact that CeRu$_2$Al$_{10}$ shows metallic characteristics~\cite{Muro2010-2}.
Ce$M_2$Al$_{10}$ is an orthorhombic YbFe$_2$Al$_{10}$ type crystal structure (space group $Cmcm$, No. 63)~\cite{Thiede1998}.
Very recently, long-range antiferromagnetic ordering with small magnetic moments has been observed below $T_0$ in CeOs$_2$Al$_{10}$ and CeRu$_2$Al$_{10}$~\cite{Adroja2010,Khalyavin2010,Robert2010}.
Since the Ce--Ce distance is about 5~\AA, which is longer than that of normal Ce compounds, however, the Ruderman-Kittel-Kasuya-Yoshida (RKKY) interaction is not believed to be the origin of the phase transition at $T_0$.
Therefore, the origin of the antiferromagnetic ordering remains unknown.

In this Letter, we describe an investigation into the origin of the anomalous phase transition at $T_0$, as well as the change in the electronic structure of CeOs$_2$Al$_{10}$, using temperature-dependent anisotropic optical conductivity [$\sigma(\omega)$] spectra.
At higher temperatures, CeOs$_2$Al$_{10}$ has the electronic structure of an anisotropic semiconductor.
Below 39~K (hereafter, $T^*$), which is higher than $T_0$, on the other hand, a pronounced peak structure of the $\sigma(\omega)$ spectra only along the $b$-axis appears at a photon energy $\hbar\omega$ of 20~meV, indicating evolution of the energy gap because of structural distortion; the $\sigma(\omega)$ spectra along the $a$- and $c$-axes are unchanged, except for a reduction in thermal broadening.
Onset of the 20-meV peak appears below $T_0$.
These observations indicate that opening of the energy gap induces antiferromagnetic ordering.


Near-normal incident polarized optical reflectivity [$R(\omega)$] spectra were acquired in a very wide photon-energy region of 2~meV -- 30~eV to ensure an accurate Kramers-Kronig analysis (KKA).
Single-crystalline CeOs$_2$Al$_{10}$ was synthesized by the Al-flux method~\cite{Muro2010-2} and was well-polished using 0.3~$\mu$m grain-size Al$_{2}$O$_{3}$ wrapping film sheets for the $R(\omega)$ measurements.
Martin-Puplett and Michelson type rapid-scan Fourier spectrometers 
(JASCO Co. Ltd., FARIS-1 and FTIR610) 
were used at photon energies $\hbar\omega$ of 2~--~30~meV and 5~meV~--~1.5~eV, respectively, with a specially designed feed-back positioning system to maintain the overall uncertainty level less than $\pm$0.5~\% at sample temperatures $T$ in the range of 8~--~300~K~\cite{Kimura2008}.
To obtain the absolute $R(\omega)$ values, the samples were evaporated {\it in-situ} with gold, whose spectrum was then measured as a reference.
At $T=300$~K, $R(\omega)$ was measured for energies 1.2--30~eV by using synchrotron radiation~\cite{Fukui2001}.
In order to obtain $\sigma(\omega)$ via KKA of $R(\omega)$, the spectra were extrapolated below 2~meV with a Hagen-Rubens function,
and above 30~eV with a free-electron approximation $R(\omega) \propto \omega^{-4}$~\cite{DG}.

%
\begin{figure}[t]
\begin{center}
\includegraphics[width=0.30\textwidth]{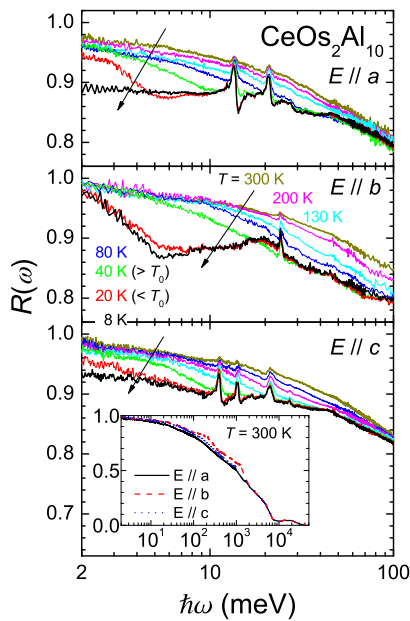}
\end{center}
\caption{
(Color online)
Low-energy portion of the temperature-dependent polarized reflectivity [$R(\omega)$] spectra of CeOs$_2$Al$_{10}$ along three principal axes.
(Inset) The entire reflectivity spectra at 300~K along the $a$-axis (solid lines), $b$-axis (dashed lines), and $c$-axis (dotted lines).
}
\label{reflectivity}
\end{figure}
The $R(\omega)$ spectra obtained along the $a$-axis ($E//a$), $b$-axis ($E//b$), and $c$-axis ($E//c$) are shown in Fig.~\ref{reflectivity}.
As can be seen in the inset, the $R(\omega)$ spectra monotonically decrease up to $\hbar\omega$~=~10~eV because the conduction band of Al expands to about 10~eV below $E_{\rm F}$ according to the band calculation (not shown).
The characteristic double-peak structure of Ce compounds in the energy range of 100--300~meV, which originates from the top of the valence band to the unoccupied Ce~$4f$ state with spin-orbit splitting~\cite{Kimura2009}, slightly appears only in $E//b$.
In addition, the X-ray photoemission spectrum of the Ce~$3d$ core level suggests mixed valence of the Ce ion (not shown)~\cite{Miyazaki2010}.
Both of these features indicate strong $c$-$f$ hybridization intensity.

Let us focus on the spectra below 100~meV.
There are two sharp peaks in $E//a$, one peak in $E//b$, and three peaks in $E//c$ in the range $\hbar\omega$~=~10--25~meV due to optical phonons. 
Except for these peaks, the $R(\omega)$ spectra for all principal axes at 300~K are Drude-like spectra that increase to unity with decreasing photon energy, indicating a metallic character.
Below 80~K, the value of $R(\omega)$ below 50~meV rapidly decreases with decreasing temperature, and eventually, at the lowest accessible photon energy, $R(\omega)$ does not close to unity at 8~K  except for $E//b$.
This indicates the existence of an energy gap; {\it i.e.}, the material changes to an insulator.

\begin{figure}[t]
\begin{center}
\includegraphics[width=0.40\textwidth]{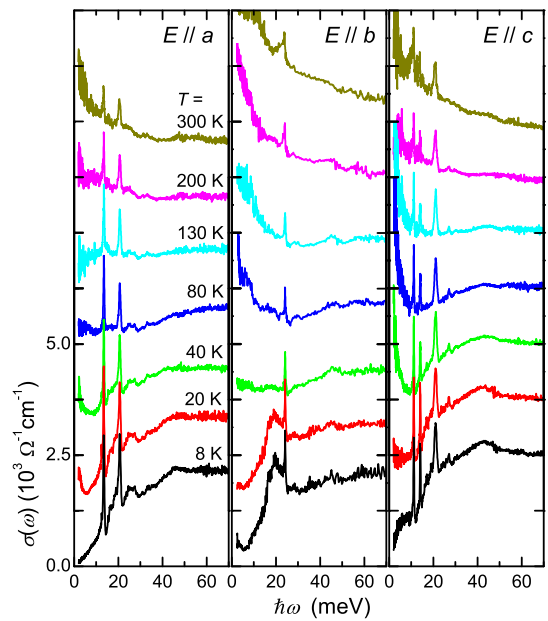}
\end{center}
\caption{
(Color online)
Temperature-dependent polarized optical conductivity [$\sigma(\omega)$] spectra of CeOs$_2$Al$_{10}$ in the photon energy region below 70~meV.
Each of the lines is shifted by 1.25 $\times10^3$~$\Omega^{-1}$cm$^{-1}$ for clarity.
}
\label{OC}
\end{figure}
The temperature-dependent $\sigma(\omega)$ spectra derived from KKA of the $R(\omega)$ spectra in Fig.~\ref{reflectivity} are shown in Fig.~\ref{OC}.
The $\sigma(\omega)$ spectra for all principal axes at 300~K monotonically increase with decreasing photon energy, again indicating a metallic character.
Commonly in all principal axes, the $\sigma(\omega)$ intensity below $\hbar\omega$~=~60~meV decreases with decreasing temperature, and at 80~K, a broad shoulder structure appears at about 45~meV, which indicates the energy gap due to the strong $c$-$f$ hybridization similar to other Kondo semiconductors~\cite{Kimura1994,Bucher1994,Okamura1998,Matsunami2003}.

In $E//a$ and $E//c$, the gentle shoulder structures at about 50~meV at 80~K gradually evolve below 130~K and an energy-gap structure appears at the low-energy side of the peak as the temperature decreases to the minimum.
However, the gap shape and energy do not change with temperature (except for the reduction in thermal broadening), unlike the case in $E//b$.
Below 40~K, a sharp Drude structure appears below 10~meV that differs from the spectral shape above 80~K.
Therefore, the gap structure drastically changes at around $T_0$.
At 8~K, the Drude structure disappears in the spectral range and a clear energy gap opens, despite the fact that the direct current conductivity along the $a$-axis is about 4000 to 3000~$\Omega^{-1}$cm$^{-1}$ in all temperature regions.
This means that a very narrow Drude peak due to in-gap states must appear below the accessible energy region.
The origin of the in-gap states is not clear at present.

\begin{figure}[t]
\begin{center}
\includegraphics[width=0.45\textwidth]{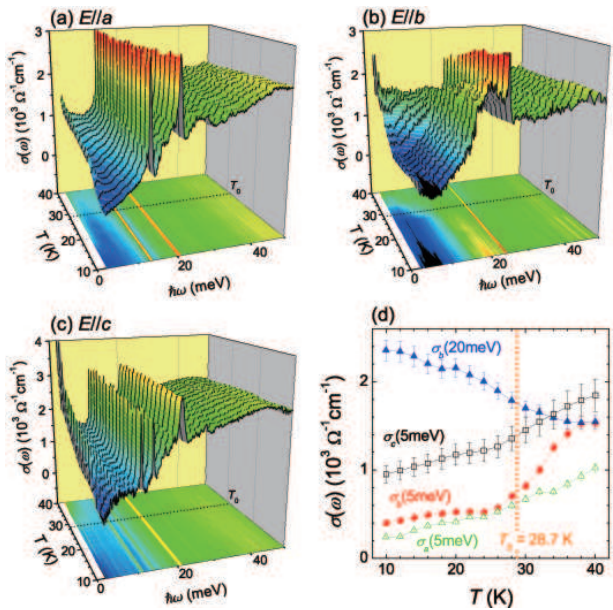}
\end{center}
\caption{
Temperature-dependent optical conductivity [$\sigma(\omega)$] spectra in $E//a$ (a), $E//b$ (b), and $E//c$ (c) at temperatures from 10 to 40~K.
(d) Temperature dependence of representatives of spectral change.
$\sigma_x({\rm 5~meV})$ [$x = a, b, c$ ($x$ is axis name)] and $\sigma_b({\rm 20~meV})$ are the intensities of the $\sigma(\omega)$ spectra at 5 and 20~meV, respectively.
}
\label{TdepOC}
\end{figure}
In $E//b$, the overall temperature dependence is similar to those in $E//a$ and $E//c$, but one peak at and an energy gap below 20~meV suddenly appear at temperatures between 20 and 40~K.
To assist in studying the spectral change in detail, Figs.~\ref{TdepOC}(a)--(c) show the fine temperature dependence ($\Delta T=$~2~K) of the $\sigma(\omega)$ spectra at temperatures from 10 to 40~K in the three principal axes.
With decreasing temperature, the spectral weight below 10~meV decreases in $E//a$ and $E//c$.
The temperature dependence of the $\sigma(\omega)$ intensity at 5~meV is plotted as $\sigma_a({\rm 5~meV})$ for $E//a$ and $\sigma_c({\rm 5~meV})$ for $E//c$ in Fig.~\ref{TdepOC}(d).
Both $\sigma_a({\rm 5~meV})$ and $\sigma_c({\rm 5~meV})$ monotonically decrease with decreasing temperature and do not show anomalies at $T_0$.
The temperature dependence is the same as that of typical Kondo semiconductors~\cite{Kimura1994,Bucher1994,Okamura1998,Matsunami2003}, {\it i.e.}, the state in $E//a$ and $E//c$ can be regarded as that of a Kondo semiconductor.

In $E//b$, on the other hand, not only does the spectral weight below 10~meV decrease, but also a peak grows at 20~meV with decreasing temperature.
As seen in the $\sigma_b({\rm 5~meV})$ and $\sigma_b({\rm 20~meV})$ behavior, the spectral weight shifts from the lower to the higher energy side across the energy gap.
The change starts at about $T^*$, not $T_0$.
$T^*$ is the temperature of maximum magnetic susceptibility ($M/H$) and of the upturn in electrical resistivity ($\rho$)~\cite{Muro2010-2}.
Therefore, the downturn in $M/H$ and upturn in $\rho$ originate from generation of the 20-meV peak.
In addition, the drastic change in the electronic structure in $E//b$ is consistent with the finding that the temperature-dependent $\rho$ in $E//b$ at around $T^*$ is the largest among those of all the principal axes.

\begin{figure}[t]
\begin{center}
\includegraphics[width=0.45\textwidth]{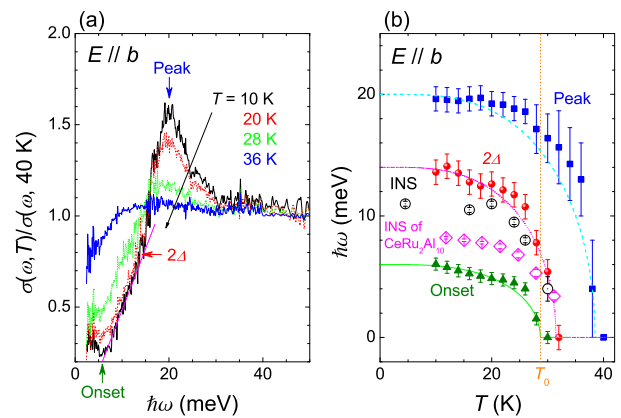}
\end{center}
\caption{
(Color online)
(a) Ratio spectra of optical conductivity [$\sigma(\omega)$] at 10, 20, 28, and 36~K, derived from that at 40~K along the $b$-axis.
(b) Temperature dependence of the peak energy (solid squares), the energy at half of the peak intensity ($2\Delta$, solid circles), and the onset (solid triangles) indicated in (a).
The inelastic neutron peaks of CeOs$_2$Al$_{10}$ (open circles, Ref.~[12]) and of CeRu$_2$Al$_{10}$ (open diamonds, Ref.~[13]) in which the temperature is normalized against $T_0$ are also plotted.
The solid, dotted, and dashed lines are the temperature-dependent energy gaps predicted by BCS theory with $T_c$~=~28.7~K ($T_0$), 31~K, and 39~K ($T^*$), respectively.
}
\label{OCdev}
\end{figure}
To clarify the difference in temperature dependence of the $\sigma(\omega)$ spectra in $E//b$, the ratio [$\sigma(\omega,T)/\sigma(\omega,{\rm 40~K})$] spectra, {\it i.e.}, $\sigma(\omega)$ at a given temperature divided by that at 40~K, are shown in Fig.~\ref{OCdev}(a).
These plots clearly demonstrate the transfer of the spectral weight across $T_0$ and $T^*$.
In the figure, a peak that appears at 20~meV at 10~K shifts to the low energy side and broadens with increasing temperature.
From the temperature-dependent peak structure, the peak energy, the energy at the half-peak intensity ($2\Delta$), and the onset of the gap are plotted as a function of temperature in Fig.~\ref{OCdev}(b).
In the same way as in Fig.~\ref{TdepOC}(d), the peak is generated below $T^*$ and does not have an anomaly at $T_0$.
However, the onset becomes visible below $T_0$.
These results mean that the electronic structure is modulated below $T^*$ and the energy gap fully opens below $T_0$.
At $T_0$, antiferromagnetic ordering is developed.
This indicates that the magnetic phase transition is driven by the electronic structure as well as the energy-gap opening in $E//b$.
The gap size $2\Delta$, indicated by the solid circles in Fig.~\ref{OCdev}(b), develops below a slightly higher temperature than $T_0$.
The temperature dependence of $2\Delta$ is consistent with the peak of inelastic neutron scattering (INS) of the same material and the related material CeRu$_2$Al$_{10}$ with the temperature normalized by $T_0$~\cite{Adroja2010,Robert2010}.
In the figure, the temperature dependence of the superconducting gap predicted by BCS theory is also plotted for reference.
The temperature dependence at all data points is similar to, but deviates slightly from, the BCS curves.
This result is consistent with the previous study of charge-density wave (CDW) and spin-density wave (SDW) transitions such as that of (TMTSF)$_2$PF$_6$~\cite{Dressel1997}.
In the case of CeOs$_2$Al$_{10}$, no magnetic transition was observed at $T^*$.
Therefore, CDW is more plausible for the origin of $T^*$.
This is consistent with the observed superlattice reflections along the $[0\bar{1}1]$ direction~\cite{Muro2010-2}.

The change in the $\sigma(\omega)$ spectrum due to the CDW and SDW transitions has previously been studied experimentally in $1T$-TiSe$_2$~\cite{Li2007}, P$_4$W$_{14}$O$_{50}$~\cite{Zhu2002}, (TMTSF)$_2$PF$_6$~\cite{Dressel1997}, Cr~\cite{Barker1968}, and others.
In all cases, the data obtained have indicated that the gap size $2\Delta$ is significantly larger than the mean-field BCS value of $3.52k_{\rm B}T_c$, where $T_c$ is the critical temperature ($\sim 5k_{\rm B}T_{c}$).
In the case of CeOs$_2$Al$_{10}$, $T_c$ can be regarded as $T^*$.
Because $2\Delta$ at 10~K is 14$\pm$1.2~meV ($\sim175\pm15$~K), the relationship between $2\Delta$ and $T^*$ becomes $2\Delta=(4.5\pm0.4)k_{\rm B}T^*$.
This is consistent with the results of previous studies.
Therefore, the CDW scenario is again supported.

%
To summarize, the electronic structure and origin of the anomalous phase transition of CeOs$_2$Al$_{10}$ were investigated by the measurement of temperature-dependent polarized optical conductivity spectra.
Along the $a$- and $c$-axes, the spectral weight at energies lower than 10~meV monotonically decreased with decreasing temperature, indicating a Kondo semiconductor characteristic with an energy gap of about 10~meV realized in all temperatures.
In contrast, along the $b$-axis, a CDW energy gap opened below 39~K, which is higher than $T_0$, and the onset of the energy gap become visible at $T_0$.
Therefore, the anomalous antiferromagnetic transition at $T_0$ is concluded to be driven by CDW transition along the $b$-axis, in contrast to the lack of phase transition along the $a$- and $c$-axes.

%
We would like to thank Mr. Hajiri for his help of Ce~$3d$ XPS measurement.
Part of this work was supported by the Use-of-UVSOR Facility Program (BL7B, 2009) of the Institute for Molecular Science.
The work was partly supported by a Grant-in-Aid for Scientific Research from MEXT of Japan (Grant No.~22340107, 20102004).

%

\begin{thebibliography}{99}
%
\bibitem{Hewson1993}
A. C. Hewson, {\it The Kondo Problem to Heavy Fermions} (Cambridge University Press, Cambridge, 1993).
%
\bibitem{Takabatake1998}
T. Takabatake, F. Iga, T. Yoshino, Y. Echizen, K. Katoh, K. Kobayashi, M. Higa, N. Shimizu, Y. Bando, G. Nakamoto, H. Fujii, K. Izawa, T. Suzuki, T. Fujita, M. Sera, M. Hiroi, K. Maezawa, S. Mock, H. v. L\"ohneysen, A. Br\"uckl, K. Neumaier, and K. Andres, J. Magn. Magn. Mater. {\bf 177-181}, 277 (1998), and references therein.
%
\bibitem{Strydom2009}
A. M. Strydom, Physica B {\bf 404}, 2981 (2009).
%
\bibitem{Nishioka2009}
T. Nishioka, Y. Kawamura, T. Takesaka, R. Kobayashi, H. Kato, M. Matsumura, K. Kodama, K. Matsubayashi, and Y. Uwatoko, J. Phys. Soc. Jpn. {\bf 78}, 123705 (2009).
%
\bibitem{Matsumura2009}
M. Matsumura, Y. Kawamura, S. Edamoto, T. Takesaka, H. Kato, T. Nishioka, Y. Tokunaga, S. Kambe, and H. Yasuoka, J. Phys. Soc. Jpn. {\bf 78}, 123713 (2009).
%
\bibitem{Muro2009}
Y. Muro, K. Motoya, Y. Saiga, and T. Takabatake, J. Phys. Soc. Jpn. {\bf 78}, 083707 (2009).
%
\bibitem{Muro2010-1}
Y. Muro, K. Motoya, Y. Saiga, and T. Takabatake, J. Phys.: Conf. Ser. {\bf 200}, 012136 (2010).
%
\bibitem{Muro2010-2}
Y. Muro, J. Kajino, K. Umeo, K. Nishimoto, R. Tamura, and T. Takabatake, Phys. Rev. B {\bf 81}, 214401 (2010).
%
\bibitem{Thiede1998}
V. M. T. Thiede, T. Ebel, and W. Jeitschko, J. Mater. Chem. {\bf 8}, 125 (1998).
%
\bibitem{Takesaka2010}
T. Takesaka, K. Oe, R. Kobayashi, Y. Kawamura, T. Nishioka, H. Kato, M. Matsumura, and K. Kodama, J. Phys.: Conf. Ser. {\bf 200}, 012201 (2010).
%
\bibitem{Khalyavin2010}
D. D. Khalyavin, A. D. Hillier, D. T. Adroja, A. M. Strydom, P. Manuel, L. C. Chapon, P. Peratheepan, K. Knight, P. Deen, C. Ritter, Y. Muro, and T. Takabatake, Phys. Rev. B {\bf 82}, 100405(R) (2010).
%
\bibitem{Adroja2010}
D. T. Adroja, A. D. Hillier, P. P. Deen, A. M. Strydom, Y. Muro, J. Kajino, W. A. Kockelmann, T. Takabatake, V.K. Anand, J.R. Stewart, and J. Taylor, Phys. Rev. B {\bf 82}, 104405 (2010).
%
\bibitem{Robert2010}
J. Robert, J. -M. Mignot, G. Andr\'e, T. Nishioka, R. Kobayashi, M. Matsumura, H. Tanida, D. Tanaka, and M. Sera, Phys. Rev. B {\bf 82}, 100404(R) (2010).
%
\bibitem{Kimura2008}
S. Kimura, JASCO Report {\bf 50}, 6 (2008). [in Japanese]
%
\bibitem{Fukui2001}
K. Fukui, H. Miura, H. Nakagawa, I. Shimoyama, K. Nakagawa, H. Okamura, T. Nanba, M. Hasumoto, and T. Kinoshita, Nucl. Instrum. Methods Phys. Res. A {\bf 467-468}, 601 (2001).
%
\bibitem{DG}
M. Dressel and G. Gr\"uner, {\it Electrodynamics of Solids} (Cambridge University Press, Cambridge, UK, 2002).
%
\bibitem{Kimura2009}
S. Kimura, T. Iizuka, and Y. S. Kwon, J. Phys. Soc. Jpn. {\bf 78}, 013710 (2009).
%
\bibitem{Miyazaki2010}
H. Miyazaki, T. Hajiri, Y. Muro, T. Takabatake, and S. Kimura, unpublished data.
%
\bibitem{Kimura1994}
S. Kimura, T. Nanba, S. Kunii, and T. Kasuya, Phys. Rev. B {\bf 50}, 1406 (1994).
%
\bibitem{Bucher1994}
B. Bucher, Z. Schlesinger, P. C. Canfield, and Z. Fisk, Phys. Rev. Lett. {\bf 72}, 522 (1994).
%
\bibitem{Okamura1998}
H. Okamura, S. Kimura, H. Shinozaki, T. Nanba, F. Iga, N. Shimizu, and T. Takabatake, Phys. Rev. B {\bf 58}, R7496 (1998).
%
\bibitem{Matsunami2003}
M. Matsunami, H. Okamura, T. Nanba, H. Sugawara, and H. Sato, J. Phys. Soc. Jpn. {\bf 72}, 2722 (2003).
%
\bibitem{Dressel1997}
M. Dressel, L. Degiorgi, J. Brickmann, A. Schwartz, and G. Gr\"uner, Physica B {\bf 230-232}, 1008 (1997).
%
\bibitem{Li2007}
G. Li, W. Z. Hu, D. Qian, D. Hsieh, M. Z. Hasan, E. Morosan, R. J. Cava, and N. L. Wang, Phys. Rev. Lett. {\bf 99}, 027404 (2007).
%
\bibitem{Zhu2002}
Z. -T. Zhu, J. L. Musfeldt, Z. S. Teweldemedhim, and M. Greenblatt, Phys. Rev. B {\bf 65}, 214519 (2002).
%
\bibitem{Barker1968}
A. S. Barker, Jr., B. I. Halperin, and T. M. Rice, Phys. Rev. Lett. {\bf 20}, 384 (1968).
%
\end{thebibliography}
\end{document}